\documentclass[pra,aps,twocolumn,showpacs,superscriptaddress]{revtex4}

\usepackage{hyperref}
\usepackage{bm}
\usepackage{epsfig}
\usepackage{amssymb}
\usepackage{graphicx}

\usepackage{amsmath}
\usepackage{epsfig}

\begin{document}
\title{Experimental test of strongly non-classical character\\ of a noisy  squeezed single-photon state}
\author{M. Je\v{z}ek}
\affiliation{Department of Optics, Palack\'{y} University, 17. listopadu 1192/12, CZ-771 46 Olomouc, Czech Republic}
\affiliation{Department of Physics, Technical University of Denmark, Fysikvej, DK-2800 Kgs. Lyngby, Denmark}

\author{A. Tipsmark}
\affiliation{Department of Physics, Technical University of Denmark, Fysikvej, DK-2800 Kgs. Lyngby, Denmark}

\author{R. Dong}
\affiliation{Department of Physics, Technical University of Denmark, Fysikvej, DK-2800 Kgs. Lyngby, Denmark}
\affiliation{Quantum Frequency Standards Division, National Time Service Center (NTSC), Chinese Academy of Sciences,
710600 Lintong, Shaanxi, China}

\author{J. Fiur\'{a}\v{s}ek}
\affiliation{Department of Optics, Palack\'{y} University, 17. listopadu 1192/12, CZ-771 46 Olomouc, Czech Republic}

\author{L. Mi\v{s}ta, Jr.}
\affiliation{Department of Optics, Palack\'{y} University, 17. listopadu 1192/12, CZ-771 46 Olomouc, Czech Republic}

\author{R. Filip}
\affiliation{Department of Optics, Palack\'{y} University, 17. listopadu 1192/12, CZ-771 46 Olomouc, Czech Republic}

\author{U.L. Andersen}
\affiliation{Department of Physics, Technical University of Denmark, Fysikvej, DK-2800 Kgs. Lyngby, Denmark}

\begin{abstract}
We experimentally verify the quantum non-Gaussian character of a conditionally generated noisy squeezed single-photon state with positive 
Wigner function. Employing an optimized witness based on probabilities of 
squeezed vacuum and squeezed single-photon states we prove that the state cannot be expressed as a mixture of Gaussian states. 
In our experiment, the non-Gaussian state is generated by conditional subtraction of a single photon from squeezed vacuum state. The state 
is probed with a homodyne detector and the witness is determined by averaging a suitable pattern function over the measured homodyne data. 
Our experimental results are in good agreement with a theoretical fit obtained from a simple yet realistic model of the experimental setup.
\end{abstract}


\maketitle

\section{Introduction}

Quantum non-classicality is a fundamental feature of quantum physics and a key resource in modern applications 
such as quantum information processing or quantum metrology. At the early stage of quantum optics, coherent states have
been identified as quantum states establishing a bridge between the classical and
quantum coherence theory \cite{Glauber}.
Coherent states can be generated from vacuum by a classical coherent driving 
described by an interaction Hamiltonian linear in the annihilation $a$ and creation $a^{\dagger}$ operators. 
Coherent states provide a quantum description of classical coherent light waves 
and their mixtures $\rho=\int P(\alpha)|\alpha\rangle\langle\alpha|\,\mathrm{d}^2\alpha$ are therefore considered to be classical
because they are described by a positive semi-definite Glauber-Sudarshan
 function $P(\alpha)$ \cite{Glauber}, which can be treated as a probability
density of wave amplitudes. On the other hand, if $P(\alpha)$ cannot be
interpreted as a probability density, the state becomes non-classical from the point
of view of coherence theory.

Operationally, non-classical states cannot be prepared using only coherent states and passive linear optical elements. 
The lowest-order nonlinearity capable of producing non-classical state is described by interaction Hamiltonians quadratic in $a$ and $a^{\dagger}$. 
Such a quadratic nonlinearity can generate non-classical squeezed states whose $P(\alpha)$ cannot be considered as an ordinary
probability distribution. However, the squeezed states can still be described by a positive semi-definite Gaussian Wigner function $W(\alpha)$ \cite{Glauber}.

In analogy to the set of mixtures of coherent states we can introduce a set $\mathcal{G}$ of all Gaussian states and their mixtures,
$\rho = \int P(\lambda) \rho_G(\lambda)\, \mathrm{d} \lambda$, where $\rho_{G}(\lambda)$ denotes a Gaussian state,
multi-index $\lambda$ labels all possible Gaussian states, and $P(\lambda)\geq 0$ is a probability density. 
All states in $\mathcal{G}$ possess positive Wigner function due to the Gaussian nature. Note that, according to Hudson theorem, 
Wigner function of any pure non-Gaussian state is negative at some points of phase space \cite{Hudson74} and these states are thus 
highly non-classical. 
On the other hand, non-Gaussian Wigner function of a mixed state $\rho$ generally does not imply that the state is highly non-classical 
as this non-Gaussianity might arise solely from a non-Gaussian distribution, $P(\lambda)$, of states. This is an example of a classical non-Gaussianity. 
In contrast we define {\it quantum} non-Gaussian states as all states which do not belong to $\mathcal{G}$ \cite{Filip11,Jezek11}. 
The quantum non-Gaussian states are highly non-classical because they cannot 
be prepared from thermal or coherent states using only Gaussian operations such as squeezing, and classical randomization. 
This means that some higher order nonlinearity is necessarily involved in the state preparation. 
In view of the no-go theorems for entanglement distillation \cite{Eisert02,Giedke02,Fiurasek02}, 
quantum error correction \cite{Niset09}, and quantum computing \cite{Bartlett02a,Bartlett02b,Ohliger10} with Gaussian states and operations, 
quantum non-Gaussian states therefore represent an enabling resource for continuous variable quantum information processing \cite{Braunstein05,Cerf07}.

It is therefore very important to develop reliable techniques for experimental identification of the quantum non-Gaussian states.
Determining whether any given state is non-classical is generally a very challenging task that may require the investigation of an 
infinite number of conditions \cite{Vogel02}. Moreover, many experiments provide only partial information about the observed state.
Fortunately, these difficulties can be overcome by formulating specific criteria that provide a sufficient condition for non-classicality.
Any such criterion can unambiguously verify non-classicality of a class of quantum states while for other states the result is inconclusive.

Recently, it has been shown that the quantum non-Gaussian character can be witnessed simply by determining the probabilities for the occurrence 
of vacuum and single-photon states $p_0=\langle 0|\rho|0\rangle$, $p_1=\langle 1|\rho|1\rangle$ \cite{Filip11}. 
This criterion allows us to conclusively certify quantum non-Gaussian character of a large set of states including those with positive
Wigner functions. Simultaneously, it can be used to detect the presence of processes with higher than quadratic quantum nonlinearity.

Detection techniques typically register either particle-like or wave-like properties of quantum states. 
In optical settings where both types of detectors can be employed, a reliable direct measurement
of the photon number is possible only for a low number of photons. When determining the photon number probabilities 
from coincidence measurements with realistic single-photon detectors that only distinguish the presence or absence of photons
we need to ensure that non-classical features are not overestimated \cite{Jezek11}. 
Alternatively, the probabilities $p_0$ and $p_1$ can be estimated indirectly by homodyne tomography \cite{Raymer93,Leonhardt97,Welsch99}. 
A homodyne detector is used to measure rotated quadratures $x_\theta=\frac{1}{\sqrt{2}}\left(a e^{-i\theta}+a^{\dagger}e^{i\theta}\right)$,
where $\theta$ is the relative phase between a strong coherent local oscillator and a signal beam, and  the probabilities
$p_0$ and $p_1$ are then reconstructed from the measured data by post-processing. 
In addition, coherent displacement or even squeezing can be applied to the experimental data
so that we can determine $p_0$ and $p_1$ of a displaced and/or squeezed version of the original state only by data processing 
without performing these operations physically on the signal mode. 
This is a very useful technique that significantly broadens the applicability of the criterion.
Highly non-classical character of many states may be masked by a Gaussian envelope which can prevent identification of quantum non-Gaussian character
from the knowledge of $p_0$ and $p_1$.  The above technique can remove this Gaussian veil and greatly enhance the power of the criterion.

In this paper we experimentally certify the quantum non-Gaussianity of a conditionally generated squeezed single-photon state
\cite{Wenger04,Ourjoumtsev06,Neergaard06,Wakui07,Takahashi08,Gerrits10,Neergaard10,Tipsmark11} 
whose Wigner function is positive at the origin of phase space due to noise. The state is prepared by conditionally 
subtracting a single photon from squeezed vacuum state and it is probed with a homodyne detector. Probabilities $p_0$ and $p_1$
are determined by two different estimation methods, namely linear reconstruction based on pattern functions \cite{D'Ariano95,Leonhardt95a,Leonhardt95b,Richter96,Richter00} 
and nonlinear maximum likelihood estimation  \cite{Hradil97,Rehacek01,Lvovsky04}.  
To optimally witness the quantum non-Gaussian character of the measured state we apply a suitable anti-squeezing operation to the experimental data. 
We conclusively prove the quantum non-Gaussian character of the prepared state with confidence of $2.4$ standard deviations. 
The experimental results are successfully fitted with a standard theoretical model of photon subtraction from squeezed vacuum.

The rest of the paper is organized as follows. The criterion allowing identification of quantum non-Gaussian states is discussed in Sec. II.
The experimental setup for the generation of photon-subtracted squeezed states is described in Sec. III and a theoretical model of the setup is presented in Sec. IV.
The reconstruction of photon number distribution from experimental data by pattern functions and maximum likelihood estimation is discussed in Sec. V. 
Experimental results are presented in Sec. VI. Finally, Sec. VII contains brief conclusions.

\section{Quantum non-Gaussian States}

A simple and powerful criterion for practical identification of quantum non-Gaussian states has been recently proposed in Ref. \cite{Filip11}.
This criterion is based on determination of the maximum probability of a single-photon state, $p_{1,G}$, that can be achieved 
by Gaussian states and their mixtures for a fixed probability of vacuum $p_{0}$. If $p_1$ exceeds this bound then the state is quantum non-Gaussian. 
In fact it suffices to maximize 
$p_1$ over pure squeezed coherent states that form extremal points of  $\mathcal{G}$. This optimization yields a 
parametric description of the dependence of $p_{1,G}$ on $p_0$ \cite{Jezek11},
\begin{equation}
p_{0}=\frac{e^{-e^r\sinh(r)}}{\cosh(r)}, \qquad
p_{1,G}=\frac{e^{4r}-1}{4}\frac{e^{-e^r\sinh(r)}}{\cosh^3(r)}, 
\label{p01NG}
\end{equation}
where  $r\in[0,\infty)$. The boundary curve (\ref{p01NG}) is plotted in Fig. 1. 
Points lying above this curve can only be obtained for states that do not belong to $\mathcal{G}$. 
Interestingly, the class of quantum non-Gaussian states is much larger
than the class of states with negative Wigner function. For example, a mixture of vacuum and single-photon states with a dominant
vacuum contribution, $\rho=p|0\rangle\langle 0|+(1-p)|1\rangle\langle 1|$, $p>\frac{1}{2}$, has a positive Wigner function 
yet it can be shown that it cannot be expressed as a mixture of Gaussian states for any $p>0$ \cite{Filip11}.

\begin{figure}[!b!]
\centerline{\includegraphics[width=\linewidth]{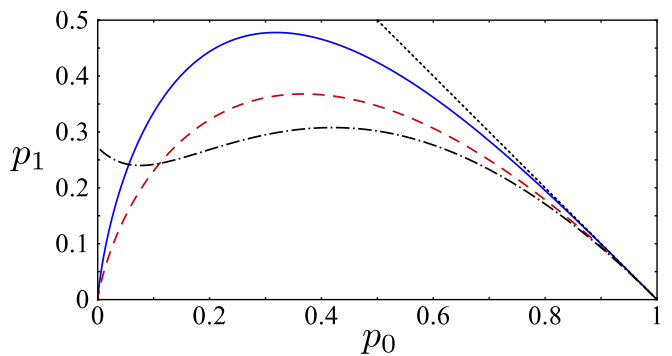}}
\caption{The maximum $p_{1}$ achievable by Gaussian states and their mixtures 
is plotted as a function of $p_0$ (solid blue line). 
The red dashed line represents the non-classicality boundary given by Eq. (\ref{p01coh}) and
the dotted black line indicates the ultimate physical boundary $p_0+p_1=1$. 
The dot-dashed black line represents a trajectory in $(p_0,p_1)$ plane of a squeezed single-photon state 
($r=1$)  subject to varying losses. The left-most point corresponds to no losses ($\eta=1$) while the right-most point 
corresponds to complete losses ($\eta=0$).}
\end{figure}

Due to the convex structure of the set $\mathcal{G}$ we can construct a witness of the state's quantum non-Gaussian character. The witness is defined
as a linear combination of $p_0$ and $p_1$,
\begin{equation}
W(a)=ap_0+p_1,
\label{Wadef}
\end{equation}
where $a<1$ is a parameter specifying the witness \cite{anote}.
The maximum value of $W(a)$ over $\mathcal{G}$ can be found by inserting formulas (\ref{p01NG}) into Eq. (\ref{Wadef}) and maximizing $W(a)$ over $r$.
After some algebra one obtains
\begin{equation}
W_G(a)=\left(a+\frac{e^{4r_0}-1}{4\cosh^2(r_0)}\right)\frac{e^{-e^r_0\sinh(r_0)}}{\cosh(r_0)},
\end{equation}
where 
\begin{equation}
 r_0=\frac{1}{2}\ln\frac{3-a+\sqrt{a^2-10a+9}}{2}
\end{equation}
is the optimal $r$ for a given $a$.
If $W(a)>W_G(a)$ then the state is quantum non-Gaussian. Each optimal witness can be represented by a straight line $ap_0+p_1=W_G(a)$ which is  
tangent to the boundary curve (\ref{p01NG}).

In a similar fashion we can also investigate whether the state is non-classical in the sense that it cannot be expressed as a mixture of coherent states 
$|\alpha\rangle$. In this case the maximum $p_1$ achievable for a fixed $p_0$ is specified by Poissonian statistics and no further optimization is necessary \cite{Lachman12},
\begin{equation}
p_{0}=e^{-\bar{n}}, \qquad p_{1}=\bar{n}e^{-\bar{n}},
\label{p01coh}
\end{equation}
where $\bar{n}\in[0,\infty)$ is the mean photon number. 
The boundary (\ref{p01coh}) is also plotted in Fig. 1 as a dashed line. 
By analogy, we can use $W(a)$ also as the non-classicality witness. The bound achievable by mixtures of coherent states reads
$W_{\mathrm{cl}}=e^{a-1}$ \cite{Lachman12}.

In this work we are interested in the quantum non-Gaussianity of approximate squeezed single-photon states obtained by photon subtraction from a squeezed vacuum 
\cite{Wenger04,Ourjoumtsev06,Neergaard06,Wakui07,Takahashi08,Gerrits10,Neergaard10,Tipsmark11}. 
In the ideal case of perfect single-photon subtraction from pure squeezed vacuum 
we would obtain pure squeezed single-photon state $|\psi(r)\rangle=S(r)|1\rangle$, where the squeezing operation reads 
\begin{equation}
S(r)=e^{-i\frac{r}{2}(xp+px)},
\end{equation}
$r$ is the squeezing constant, and $x$ and $p$ denote conjugate quadrature operators satisfying canonical commutation relations, $[x,p]=i$.  
The witness $W(a)$ identifies the pure state $|\psi(r)\rangle$ as quantum non-Gaussian for any amount of squeezing, because it is a superposition 
of odd Fock states and $p_0=0$ while $p_1>0$. As a more realistic case let us next consider a mixed state obtained by sending $|\psi(r)\rangle$
through a lossy channel $\mathcal{L}$ with transmittance $\eta$. In Fig. 1 we plot the trajectory of points $p_0,p_1$ generated by varying the transmittance
$\eta$ for a fixed amount of initial squeezing $r$. We can see that the curve enters the region of Gaussian mixtures so for high enough losses we can no longer
identify the state as quantum non-Gaussian or as non-classical. There exists a threshold transmittance $\eta_{\mathrm{th}}$ 
for which our criterion reveals the quantum non-Gaussian character of the  state. The dependence of $\eta_{\mathrm{th}}$ on the initial squeezing $r$ is plotted in Fig. 2(a).

\begin{figure}[!t!]
\centerline{\includegraphics[width=\linewidth]{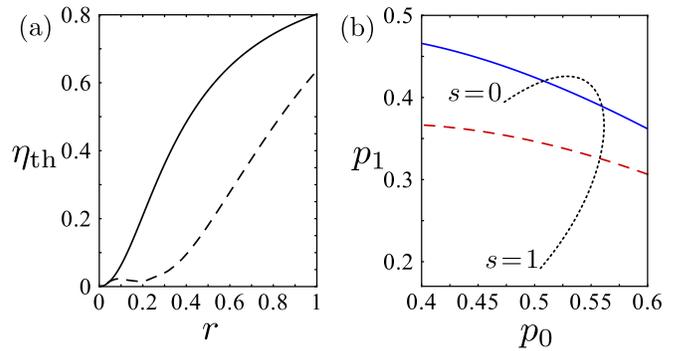}}
\caption{(a) Threshold transmittances $\eta_{\mathrm{th}}$ (solid line) and $\eta_{\mathrm{th},s}$ (dashed line) 
corresponding to witnesses $W(a)$ and $W(a,s)$, respectively, are plotted as function of squeezing constant. 
(b) A $(p_0,p_1)$ plane trajectory of a lossy squeezed single-photon state ($r=0.5$, $\eta=0.4$) subject 
to anti-squeezing operation with a variable degree of anti-squeezing $s$ (dotted black line).
The blue solid line and red dashed line have the same meaning as in Fig.~1. }
\end{figure}

Since we deal with a squeezed state, we can intuitively expect that a witness based on Fock state probabilities will not be optimal. We can significantly 
improve the performance of our witness if we first extract the highly non-classical core of the state \cite{Menzies09} by 
anti-squeezing the state, $\rho_S= S^\dagger(s) \rho S(s)$. 
Note that the squeezing is a Gaussian operation that maps mixtures of Gaussian states onto mixtures of Gaussian states. 
Therefore, if $\rho_S$ is quantum non-Gaussian then $\rho$ is also quantum non-Gaussian.
Equivalently, we can introduce a generalized witness 
\begin{equation}
W(a,s)=ap_0(s)+p_1(s),
\end{equation}
where $p_n(s)=\langle n|S^\dagger(s)\rho S(s) |n\rangle$ 
are diagonal density matrix elements in basis of squeezed Fock states $S(s)|n\rangle$. 

The anti-squeezing partly compensates for the initial squeezing
and makes the state closer to a mixture of single-photon and vacuum states for which the witness (\ref{Wadef}) is very powerful. 
Note that since the lossy channel and squeezing do not commute, $S\mathcal{L}(\rho)S^\dagger \neq \mathcal{L}(S \rho S^\dagger)$, 
the anti-squeezing of the output mixed state is not fully equivalent to reducing the initial squeezing $r$. In Fig. 2(b) we plot a typical trajectory
of $p_0,p_1$ pairs when the anti-squeezing $s$ is varied. We can see that for a suitably chosen $s$ we can detect the quantum non-Gaussian character of the state
although without anti-squeezing (corresponding to $s=0$) our witness fails. For comparison we plot in Fig. 2(a) 
the threshold transmittance $\eta_{\mathrm{th},s}$ above which the witness $W(a,s)$
detects quantum non-Gaussian character of the considered state. We can see that $\eta_{\mathrm{th},s}<\eta_{\mathrm{th}}$ 
and this gap is a clear indication of the enhanced power and increased applicability
of the witness $W(a,s)$. Note that in addition to anti-squeezing we could also coherently displace the state to remove any coherent component. 
However, this is not necessary for our present purpose because the states studied in this paper do not contain any coherent component.

\begin{figure}[!b!]
\centerline{\includegraphics[width=1.0\linewidth]{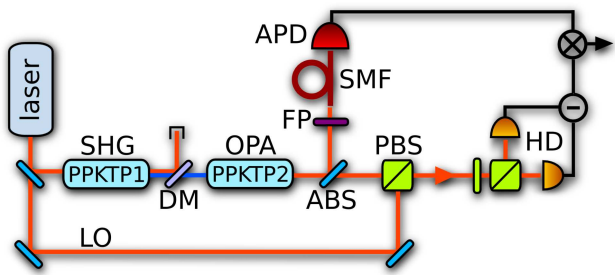}}
\caption{Experimental set-up used to generate photon-subtracted squeezed states.}
\label{catsetupfig}
\end{figure}

\section{Experimental setup}

The set-up used for the experimental test of non-classical character
of photon-subtracted squeezed states \cite{Tipsmark11} is shown in Fig.~\ref{catsetupfig}.
A cavity-dumped Ti:sapphire laser (Tiger-PS, Time-Bandwidth Products) produces
5 ps long pulses with a maximum energy of 50 nJ, a repetition rate of 815 kHz and
a central wavelength of 830 nm.
The laser pulses are up-converted in the process of second harmonic generation (SHG)
using a 3 mm thick periodically poled crystal of potassium titanyl phosphate (PPKTP1).
The remaining radiation at 830 nm is removed by a set of dichroic mirrors (DM).
The frequency doubled pulses are used as a pump in a similar crystal (PPKTP2), phase matched for colinear
and fully degenerate parametric generation (OPA). It produces a squeezed vacuum with squeezing strength
tunable from 0 to 3.5~dB. 

The generated squeezed light impinges onto an asymmetric beam splitter (ABS) which reflects $7{.}7$\%
of the signal to a single-photon detection setup. It consists of a narrowband Fabry-Perot
filter (FP, FWHM$=\!\!0.04$\,nm) and a single mode fiber (SMF) that guides the signal
to an avalanche photo diode (APD) operated in Geiger mode (SPCM-AQR-14, Perkin-Elmer).
We estimate the total quantum efficiency of the filtering and subsequent detection
process to be approximately 8$\pm$1\%. The uncertainty of the efficiency is quite high
particularly due to the uncertainty of the detection probability of the APD.
The signal transmitted by the ABS is mixed with a local oscillator (LO) using polarizing beam splitters (PBS1, PBS2) and
a half-wave plate (HWP), and detected by means of a pair of photo diodes (S3883,
Hamamatsu). The difference of the photo currents is generated and
the resulting current is amplified by a charge-sensitive amplifier
and finally fed into an oscilloscope working in the memory-segmentation regime.
The acquisition is triggered by a detection event of APD and a synchronization
pulse from the laser which suppresses the effect of electronic dark counts
(effectively below 3 Hz). The homodyne detector (HD) efficiency of 80$\pm$3\%
is limited mainly by the efficiency of the photo diodes itself (94\%),
the mode matching of signal and LO (95\%), and the transmittance of passive
optical elements in the signal path (95\%).

\section{Theoretical model}

\begin{figure}[!b!]
\centerline{\includegraphics[width=0.7\linewidth]{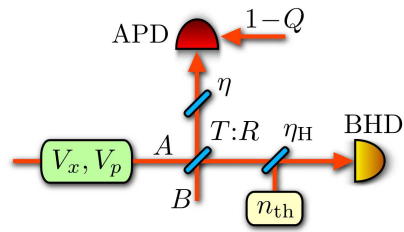}}
\caption{Equivalent theoretical model of the experimental setup.  $V_{x}$ and $V_p$ denote variances of amplitude 
and phase quadratures of the input mixed squeezed vacuum state, $T$ is the transmittance of tap-off beam splitter. 
Inefficient single-photon (homodyne) detection is modeled
as a sequence of a beam splitter with transmittance $\eta$ ($\eta_H$) followed by perfect detector. Thermal photons with mean number $n_{\mathrm{th}}$ 
can be injected into the signal beam which models electronic noise in homodyne detector BHD. Single-photon detector APD
can be triggered by dark counts or photons not coming from the signal mode which occurs with probability $1-Q$.
}
\end{figure}
To compare the experimental results with theoretical predictions we employ a standard model of squeezed 
single-photon state preparation \cite{Fiurasek05,Sasaki06,Brouri09}. 
An equivalent scheme of the experimental setup is shown in Fig. 4 and our goal is to determine 
the probabilities $p_0$ and $p_1$ as functions of model parameters $V_x$, $V_p$, $T$, $\eta$, $\eta_H$, $n_{\mathrm{th}}$, and $Q$. 
Our derivation is based on phase-space  representation and Gaussian-state formalism because
the non-Gaussian quantum state prepared by photon subtraction can be expressed 
as a weighted difference of two Gaussian states \cite{Fiurasek05}. 

The input signal mode $A_{\mathrm{in}}$ is prepared in a generally mixed
squeezed vacuum state $\rho_{\mathrm{in}}$ described by a diagonal covariance matrix (CM) $\gamma_{A,\mathrm{in}}=\mbox{diag}\left(2V_{x},2V_{p}\right)$, 
where 
\begin{equation}
V_{x}=\langle\left(\Delta x_{A_{\mathrm{in}}}\right)^{2}\rangle,\quad V_{p}=\langle\left(\Delta p_{A_\mathrm{in}}\right)^{2}\rangle 
\end{equation}
denote the variances of squeezed and anti-squeezed quadratures, respectively, and $V_{x}V_{p}\geq \frac{1}{4}$. 
A small fraction of light is then tapped-off by a strongly unbalanced beam splitter BS with intensity transmittance $T$ and reflectance $R=1-T \ll 1$.
We assume that the auxiliary input port B of BS is in the vacuum state. The state of modes $A$ and $B$ at the output of beam 
splitter is Gaussian \cite{Weebrook12} and its Wigner function reads,
\begin{eqnarray}\label{Wina}
W_{AB}(\xi)=\frac{1}{\pi^2\sqrt{\det\gamma_{AB}}} e^{- \xi^{T} \gamma_{AB}^{-1}\xi}.
\end{eqnarray}
Here $\xi=\left(x_A,p_A,x_B,p_B\right)^{T}$ is the phase space 
coordinate vector and the two-mode covariance matrix $\gamma_{AB}$ is given by
\begin{equation}\label{gamma}
\gamma_{AB}=\left( \begin{array}{cc}
T\gamma_{{A_\mathrm{in}}}+R\gamma_{{B_\mathrm{in}}} & \sqrt{RT}(\gamma_{{A_\mathrm{in}}}-\gamma_{{B_\mathrm{in}}}) \\
\sqrt{RT}(\gamma_{{A_\mathrm{in}}}-\gamma_{{B_\mathrm{in}}}) & R\gamma_{{A_\mathrm{in}}}+T\gamma_{{B_\mathrm{in}}}
\end{array}
\right),
\end{equation}
where $\gamma_{B_\mathrm{in}}=\openone$ is the covariance matrix of vacuum.
It is convenient to model inefficient single-photon detection as a combination of a lossy channel with transmittance $\eta$ followed by a
perfect detector with unit efficiency. Similarly we can model homodyne detection with efficiency $\eta_{H}$ and background noise
$n_{\mathrm{th}}$ by a sequence of a lossy channel with added thermal noise followed by a perfect homodyne detector.
The two-mode covariance matrix that accounts for detection inefficiency and noise can be expressed as follows,
\begin{equation}
\gamma_{AB}'=M\gamma_{AB}M^T+G,
\end{equation}
where 
\begin{equation}
M=\left(
\begin{array}{cccc}
\sqrt{\eta_{H}} & 0 & 0  & 0 \\
 0 & \sqrt{\eta_{H}} & 0  & 0 \\
 0 & 0 & \sqrt{\eta}  & 0  \\
 0 & 0 & 0  & \sqrt{\eta} 
\end{array}
\right),
\end{equation}
and
\begin{equation}
G=\left(
\begin{array}{cccc}
1-\eta_{H}+2n_{\mathrm{th}} & 0 & 0  & 0 \\
 0 & 1-\eta_{H}+2n_{\mathrm{th}} & 0  & 0 \\
 0 & 0 & 1-\eta  & 0  \\
 0 & 0 & 0  & 1-\eta
\end{array}
\right).
\end{equation}
The on/off detector placed on mode $B$ can be described  by  projectors onto vacuum and the rest of the Hilbert space, respectively, 
$\Pi_{0}=|0\rangle\langle 0|$ (no click) and  $\Pi_{1}=\openone -\Pi_0$ (click).
If a click of the detector occurs the state of  mode $A$ collapses into the output state
\begin{equation}\label{rhoout}
\rho_{\mathrm{out}}=\frac{\mbox{Tr}_{B}\left(\openone_A\otimes\Pi_{1,B}\rho_{AB}\right)}{\mbox{Tr}\left(\openone_A\otimes\Pi_{1,B}\rho_{AB}\right)}=
\frac{1}{1-{\cal P}_{0}}\left(\rho^{(\mathrm{red})}-{\cal P}_{0}\rho^{(0)}\right),
\end{equation}
where $\rho_{AB}$ is the joint state of modes $A$ and $B$ in front of the detectors,
$\rho^{(\mathrm{red})}=\mbox{Tr}_{B}\left(\rho_{AB}\right)$ is the reduced state of the output mode $A$, 
${\cal P}_{0}=\mbox{Tr}\left(\openone_A\otimes \Pi_{0,B}\,\rho_{AB}\right)$ is the probability of no click, and
$\rho^{(0)}=\mbox{Tr}_{A}\left(\openone_A\otimes\Pi_{0,B}\,\rho_{AB}\right)/{\cal P}_{0}$ is the conditional state of  mode $A$
corresponding to no click of the detector. 

The output state can be most easily calculated using the formalism of Wigner functions bacause the POVM element $\Pi_0$ is a projector onto vacuum that 
possesses a Gaussian Wigner function. Wigner function of the output state (\ref{rhoout}) can be written as a difference of two Gaussian Wigner functions
centered on origin,
\begin{equation}\label{Wout}
W_{\rm out}(\xi_A)=\frac{1}{\pi(1-{\cal P}_{0})}\left[
\frac{ e^{-\xi_A^T\gamma_{\mathrm{I}}^{-1}\xi_A}}{\sqrt{\det\gamma_{\mathrm{I}}}}
-\mathcal{P}_0\frac{ e^{-\xi_A^T\gamma_{0}^{-1}\xi_A}}{\sqrt{\det\gamma_{0}}}
 \right].
\end{equation}
Here $\xi_A=(x_A,p_A)^T$. In order to express the covariance matrices appearing in Eq. (\ref{Wout}) we split the 
two-mode covariance matrix $\gamma_{AB}'$ into single-mode blocks,
\begin{equation}
\gamma_{AB}'=\left(
\begin{array}{cc}
\Gamma_A & \Gamma_C \\
\Gamma_C^T & \Gamma_B 
\end{array}
\right)
\end{equation}
The covariance matrices read $\gamma_{I}=\Gamma_A$ and $\gamma_{0}=\Gamma_A-\Gamma_C(\Gamma_B+\openone)^{-1}\Gamma_C^T$ \cite{Giedke02} and for the 
 probability of no click we have,
\begin{equation}\label{calP0}
{\cal P}_{0}=\frac{2}{\sqrt{\det(\Gamma_B+\openone)}}.
\end{equation}

Despite heavy spatial and spectral filtering, the single photon detector APD can be sometimes triggered by photons from other modes than the mode
which is subsequently observed by homodyne detector. Such false triggers can also include dark counts caused by thermal fluctuations or other effects. 
We can define an effective mode overlap $Q$ as a probability that a click from the APD is caused by
photons subtracted from the right mode.  If the APD is triggered by a photon coming
from some other mode then the output state in mode A is prepared in a Gaussian state $\rho^{(\mathrm{red})}$ with zero mean and covariance matrix $\gamma_{I}$. The overall state in mode
A is then a mixture of state (\ref{rhoout}) with probability $Q$  and a Gaussian state $\rho^{(\mathrm{red})}$ with probability $1-Q$. The Wigner function of the resulting state preserves 
the form (\ref{Wout}), only $\mathcal{P}_0$ is replaced with
\begin{equation}
\mathcal{P}_0^\prime=\frac{Q\mathcal{P}_0}{1-\mathcal{P}_0(1-Q )}.
\end{equation}
Anti-squeezing operation on the output state transforms the covariance matrix of each constituent Gaussian component according to
$\gamma_j \rightarrow S\gamma_j S^T$, where $S=\mathrm{diag}(e^{-s},e^{s})$.

In order to evaluate the witness $W(a)$ we need to calculate the probabilities $p_0$ and
$p_1$ of finding the output state (\ref{Wout}) in the vacuum state and single-photon Fock
state, respectively. The probabilities can be derived from the overlap formula
\begin{eqnarray}\label{pk}
p_{k}=2\pi\int_{-\infty}^{\infty}\int_{-\infty}^{\infty} W_{\mathrm{out}}\left(\xi_A\right)W_{|k\rangle}(\xi_A)\,\mathrm{d}^2\xi_A,
\end{eqnarray}
where 
\begin{eqnarray}\label{Wk}
W_{|k\rangle}(\xi_A)=\frac{1}{\pi}\left[2k\xi_A^{T}\xi_A+(-1)^k\right]\mbox{exp}\left(-\xi_A^{T}\xi_A\right),
\end{eqnarray}
are Wigner functions of vacuum ($k=0$) and single-photon Fock state ($k=1$), respectively.
Performing integration in Eq.~(\ref{pk}) we arrive at the probabilities $p_{0}$ and $p_{1}$
in the following form,
\begin{eqnarray}\label{p0p1}
p_{0}&=&\frac{2}{1-{\cal P}_{0}'}\left[\frac{1}{\sqrt{\det\left(\gamma_{I}+\openone\right)}}-\frac{{\cal P}_{0}'}{\sqrt{\det\left(\gamma_{0}+\openone\right)}}\right],\nonumber \\
p_{1}&=&\frac{2}{1-{\cal P}_{0}'}\left\{\frac{\det\gamma_{I}-1}{\left[\det\left(\gamma_{I}+\openone\right)\right]^{\frac{3}{2}}}
-\frac{{\cal P}_{0}'\left(\det\gamma_{0}-1\right)}{\left[\det\left(\gamma_{0}+\openone\right)\right]^{\frac{3}{2}}}\right\},\nonumber
\end{eqnarray}
In Sec. VI we will use these formulas to find the best theoretical fit to the experimental data.

\section{Witness estimation}

In this section we describe the estimation techniques that were used for determination of the probabilities $p_0(s)$ and $p_1(s)$ from the experimental data. 
As a main tool we utilize the pattern functions that provide unbiased linear estimators of $p_j(s)$ and allow for straightforward estimation of error bars, which is particularly important in the present case.
For the sake of comparison we also perform maximum likelihood estimation of $p_n(s)$.

\subsection{Pattern functions}

\begin{figure}[!b!]
\centerline{\includegraphics[width=0.95\linewidth]{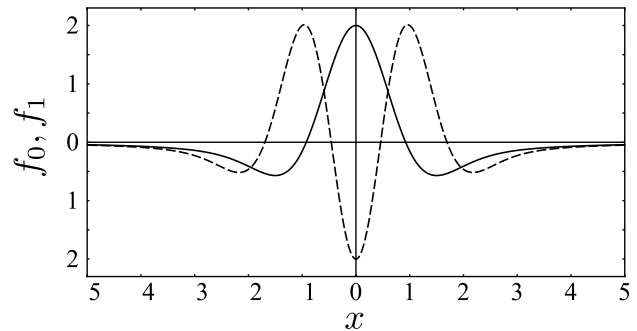}}
\caption{Pattern functions $f_0(x)$ (solid line) and $f_1(x)$ (dashed line).}
\label{f01figure}
\end{figure}

A homodyne detector measures the probability distribution $w(x_\theta;\theta)$ of a rotated quadrature operator $x_\theta=x\cos\theta+p\sin \theta$  where $\theta$ is the relative
phase between local oscillator and signal beam \cite{Leonhardt97,Welsch99}. If the measurement is performed for values of $\theta$ spanning the whole interval $\theta\in[0,\pi]$
then the measurement is tomographically complete and any element of the density matrix $\rho$ can be determined from the recorded data. 
We are interested in estimation of diagonal density matrix elements in Fock basis, i.e. photon number probabilities $p_0$ and $p_1$. A particularly straightforward method 
is to determine estimates of $p_n$ by averaging appropriate pattern functions $f_n(x_\theta)$ over the sampled quadrature statistics \cite{D'Ariano95,Leonhardt95a,Leonhardt95b,Richter96,Richter00},
\begin{equation}
p_{n}=\frac{1}{\pi}  \int_{0}^\pi \int_{-\infty}^\infty  w(x_\theta;\theta) f_n(x_\theta) \,\mathrm{d}x_\theta \,\mathrm{d}\theta .
\label{pnintegral}
\end{equation}
Note that the pattern functions $f_n$ do not depend on the phase $\theta$. For vacuum and single-photon probabilities we explicitly have \cite{Leonhardt95a,Leonhardt95b,Richter96}
\begin{eqnarray}
f_0(x)&=&2 - 2  \sqrt{\pi} x e^{-x^2} \mathrm{erfi}(x),\nonumber \\
f_1(x)&=&2(2x^2-1) + 4 \sqrt{\pi} x (1-x^2)e^{-x^2} \mathrm{erfi}(x), \nonumber \\
\label{f01simplest}
\end{eqnarray}
where $\mathrm{erfi}(x)$ denotes error function of imaginary argument. The functions (\ref{f01simplest}) are plotted in Fig. \ref{f01figure} 
and we can see that they are bounded and asymptotically approach $0$
in the limit of large $|x|$. The pattern function for estimation of witness $W(a)$ can be obtained as an appropriate linear 
combination of the two functions (\ref{f01simplest}), $f_W=af_0+f_1$.

Direct reconstruction based on pattern functions is an appealing option in situations where we need to determine just some particular
property of the state such as the witness $W(a)$ and do not want to reconstruct the whole state. In comparison with more sophisticated statistical 
reconstruction methods such as maximum likelihood estimation, the linear estimation is much faster and  the statistical uncertainty of any estimated 
quantity can be easily determined \cite{D'Ariano99} without calculating the Fisher matrix or performing complicated Monte Carlo simulations. To see this let us assume 
that the quadratures were measured for $K$ different values of the phase shift $\theta_k=\frac{k}{K}\pi$, $k=1,\ldots,K$. In the actual experiment, the phase is approximately 
linearly modulated in time and 
is then fitted and binned into $K=40$ discrete equidistant values in the $[0,\pi]$ interval, so the following treatment is applicable to our data.
Let $M_k$ denote the number of quadrature measurements for phase $\theta_k$ and let $X_{k,m}$ denote measurement outcomes 
for this setting, $m=1,\ldots, M_k$. In experimental data processing, the integral (\ref{pnintegral}) is replaced with a finite sum,
\begin{equation}
p_{n}=\frac{1}{K}\sum_{k=1}^{K} \frac{1}{M_{k}}\sum_{m=1}^{M_{k}}  f_n(X_{k,m}). 
\label{pnsum}
\end{equation}
The inner sum represents averaging of the pattern function $f_{n}$ over sampled quadrature statistics for a fixed $\theta_k$, while the outer sum
represents averaging over the phase $\theta$. If the number of quadrature samples $M_{k}$ is not a constant then measurement results
obtained for different values of $\theta$ have different weight in Eq. (\ref{pnsum}). 

Statistical uncertainty of estimated $p_n$ can be quantified by its variance $V(p_n)=\langle p_n^2\rangle-\langle p_n\rangle^2$.
Since $X_{k,m}$ are independent random variables, we find that
\begin{equation}
V(p_n)=\frac{1}{K^2} \sum_{k=1}^K V_k (f_n),
\label{Vpn}
\end{equation}
where
\begin{equation}
V_k (f_n)= \frac{1}{M_k^2} \sum_{m=1}^{M_k}(\langle f_n^2(X_{k,m}) \rangle-\langle f_n(X_{k,m}) \rangle^2).
\label{Vkfndefinition}
\end{equation}
All quadrature measurement outcomes $X_{k,m}$ obtained for a fixed $k$ exhibit the same statistical distribution, therefore 
the statistical averages appearing in  (\ref{Vkfndefinition}) can be estimated from the measured data and we obtain
\begin{equation}
V_k (f_n)= \frac{1}{M_k^2}\sum_{m=1}^{M_k}f_n^2(X_{k,m})-\frac{1}{M_k^3}\left[\sum_{m=1}^{M_k}f_n(X_{k,m})\right]^2.
\end{equation}
It follows from Eqs. (\ref{Vpn}) and (\ref{Vkfndefinition}) that the variance $V(p_n)$ scales as $1/N$, where $N$ is the total number of measurements.

\subsection{Data anti-squeezing} 
As we have seen in Sec. II, the efficiency of our witness can be greatly increased if we construct the witness from probabilities of squeezed Fock states.
Equivalently, this means that we should anti-squeeze the state before we estimate the probabilities $p_0$ and $p_1$.
In the Heisenberg picture the anti-squeezing transformation boils down to the linear re-scaling of quadratures,
\begin{equation}
x=x_0 e^{s}, \qquad p=p_0 e^{-s}.
\end{equation}
This suggests that it may be possible to perform the anti-squeezing on the homodyne data without altering the experimental setup.
In order to show this we express the measured quadrature operator $x_\theta=x\cos\theta+p\sin\theta$ in terms of the anti-squeezed quadratures $x_0$ and $p_0$,
\begin{equation}
x_{\theta}= x_0 e^s \cos\theta+p_0 e^{-s} \sin\theta.
\end{equation}
We can rewrite this expression as follows,
\begin{equation}
x_{\theta}= g(x_0 \cos\vartheta + p_0 \sin\vartheta)=g\tilde{x}_\vartheta,
\label{xvartheta}
\end{equation}
where the new effective phase $\vartheta$ and scaling factor $g$ are given by
\begin{equation}
\tan\vartheta=e^{-2s}\tan\theta, 
\label{vartheta}
\end{equation}
\begin{equation}
g=\sqrt{e^{2s}\cos^2\theta+e^{-2s}\sin^2\theta}.
\label{a}
\end{equation}
According to the above formulas, the measurement of $x_{\theta}$ can be equivalently interpreted as
measurement of the quadrature $\tilde{x}_\vartheta$ of the anti-squeezed state, where $\tilde{x}_\vartheta=x_\theta/g$. 
Photon number distribution $p_{n}(s)$ of the anti-squeezed state can be inferred by a modified formula (\ref{pnintegral}),
\begin{equation}
p_{n}(s)=\frac{1}{\pi}  \int_{0}^\pi \int_{-\infty}^\infty  \frac{1}{g^2}
w(x_\theta;\theta) f_n\left(\frac{x_\theta}{g}\right) \,\mathrm{d}x_\theta \,\mathrm{d}\theta .
\label{pnsqueezed}
\end{equation}
The factor 
\begin{equation}
\frac{1}{g^2}= \frac{\mathrm{d}\vartheta}{\mathrm{d}\theta}
\end{equation}
attributes different weights to the pattern function according to the value of $\theta$ because a homogeneous sampling over $\theta$
is equivalent to an inhomogeneous sampling over $\vartheta$ due to the nontrivial dependence of $\vartheta$ on $\theta$. Looking more carefully 
at Eq. (\ref{pnsqueezed}) we can identify the effective pattern functions for reconstruction of diagonal density matrix elements 
in the basis of squeezed Fock states,
\begin{equation}
f_{n}(x_\theta,\theta;s)=\frac{1}{g^2}f_n\left(\frac{x_\theta}{g}\right).
\label{fnsqueezed}
\end{equation}
Note that these generalized pattern functions explicitly depend on $\theta$ through $g$.

\subsection{Maximum likelihood estimation}

For comparison, we also perform maximum likelihood estimation \cite{Hradil97,Rehacek01,Lvovsky04} of the photon number distribution $p_n$. 
For this purpose we construct a quadrature histogram corresponding to averaged quadrature statistics 
\begin{equation}
\bar{w}(x)=\frac{1}{\pi}\int_0^\pi w(x;\theta) d\theta. 
\end{equation}
We divide the real axis into equidistant bins with width $\Delta x$.
To each measured value $X_{k,m}$ we assign an index $j$ of the corresponding quadrature bin, $j=\mathrm{round}(X_{k,m}/\Delta x)$,
and increase the counter for this bin by $1/M_k$, $C_j\rightarrow C_j+1/M_k$, where initially $C_j=0$. The resulting values $C_j$ 
are proportional to the probabilities that the quadrature outcome falls into the $j$th window, $P_j=\int_{(j-\frac{1}{2})\Delta x}^{(j+\frac{1}{2})\Delta x}\bar{w}(x)\,\mathrm{d} x.$
We can associate a POVM element $\Pi_j$ with each bin, $P_j=\mathrm{Tr}[\Pi_j \rho]$. Due to the phase averaging the POVM elements are 
diagonal in Fock state basis, $\Pi_j=\sum_{n=0}^\infty \Pi_{j,n}|n\rangle\langle n|$, $\Pi_{j,n} \geq 0$, and $\sum_{j}\Pi_{j,n}=1$. 
The probability of obtaining outcome in $j$th bin is given by $P_{j}=\sum_{n}\Pi_{j,n}p_n$ and  the likelihood function reads 
\begin{equation}
L=\prod_j \left(\sum_n p_n \Pi_{j,n}\right)^{C_j}.
\end{equation}
The maximum likelihood estimates $p_n$ which maximize the likelihood function $L$ can be numerically determined  by an expectation-maximization algorithm \cite{Dempster77,Vardi93,Rehacek01}.

We can generalize the above procedure to estimation of $p_n$ of the anti-squeezed states. 
According to the formula (\ref{xvartheta}), we need to properly re-scale each measurement outcome,
hence the bin index is now given by $j=\mathrm{round}[X_{k,m}/(g \Delta x)]$. Also we must take into account the weight factor 
$\frac{\mathrm{d}\vartheta}{\mathrm{d}\theta} = g^{-2}$ when building the phase-averaged histogram,
$C_{j}\rightarrow C_j+1/(g^2 M_k)$. In this way we obtain a quadrature histogram that corresponds to the phase-averaged quadrature statistics of the anti-squeezed state.
Once the histogram is calculated the probabilities $p_n(s)$ can be determined by the expectation-maximization algorithm similarly as before.

\section{Results and discussion}

\begin{figure}[!t!]
\centerline{\includegraphics[width=\linewidth]{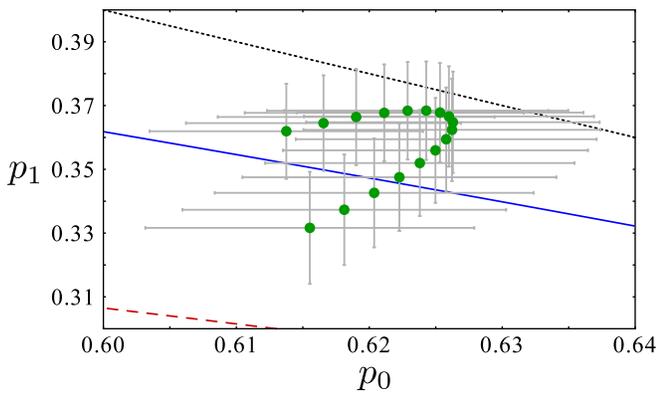}}
\caption{Probability pairs  $p_0(s)$ and $p_1(s)$ estimated from experimental data with the use of pattern functions (green circles). Error bars indicate statistical 
errors (one standard deviation). Blue solid line represents the boundary above which the state is recognized as quantum non-Gaussian. Red dashed line indicates the non-classicality boundary and 
dotted black line is the ultimate physical boundary $p_0+p_1=1$.}
\label{resultsfig}
\end{figure}

We have estimated the probabilities $p_0(s)$ and $p_1(s)$ from the experimental data using the pattern functions described in the previous section. The results are
shown in Fig.~\ref{resultsfig}, where the probability pairs $p_0(s),p_1(s)$ are depicted for $s\in[0,0.4]$. The graph suggests that the state 
is both nonclassical and quantum non-Gaussian. However, the statistical errors are significant due to a relatively low number of quadrature samples 
$N=8000$ which leads  to standard deviations of the order of $1/\sqrt{N}\approx 0.011$.  
 
\begin{figure}[!t!]
\centerline{\includegraphics[width=\linewidth]{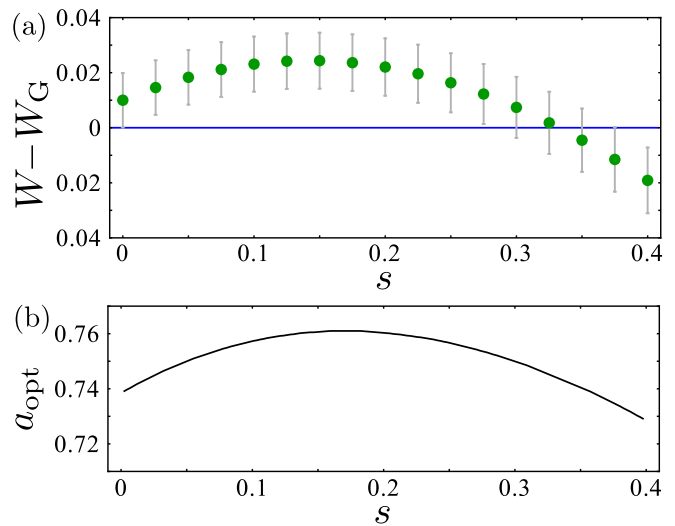}}
\caption{(a) The optimal witness $W(a_{\mathrm{opt}},s)-W_G(a_{\mathrm{opt}})$ and (b) the optimal witness parameter $a_{\mathrm{opt}}$
are plotted as functions of the anti-squeezing parameter $s$. The error bars represent one standard deviation.}
\label{witnessfig}
\end{figure}

Since there could be statistical correlations between estimates of $p_0(s)$ and $p_1(s)$, a proper evaluation of the statistical significance 
of the observed non-Gaussian character requires determination of statistical error of the witness $W(a,s)$. This is accomplished by calculating the statistical error
for the sampling function $f_W$ as discussed in Sec. V.A.  We introduce a normalized relative witness 
\begin{equation}
W_R(a,s)=\frac{W(a,s)-W_G(a)}{\Delta W(a,s)},
\end{equation}
where $\Delta W(a,s)$ is the standard deviation of $W(a,s)$. For each anti-squeezing $s$ we maximize $W_R$ over $a$. The resulting optimal 
witnesses are plotted in Fig. \ref{witnessfig}(a) and the dependence of the optimal value of $a$ on $s$ is shown in Fig. \ref{witnessfig}(b). 
Without anti-squeezing the confirmation of non-Gaussian character is not statistically significant but for optimal value of anti-squeezing 
the quantum non-Gaussian character is confirmed with much higher confidence.  The optimum squeezing yielding maximum relative witness 
is $s_{\mathrm{opt}}=0.15$ and we have 
\[
W(a_{\mathrm{opt}},s_{\mathrm{opt}})-W_G(a_{\mathrm{opt}})=0.024 \pm 0.010,
\]
hence the Gaussian threshold is exceeded by $2.4$ standard deviations. Since $p_0>\frac{1}{2}$, the Wigner function of the state is positive at the origin, where we expect it to be most negative
for this kind of state. The strongly non-classical character of the state indicated by the witness thus could not be uncovered by looking at the negativity of Wigner function.

\begin{figure*}[!t!]
\centerline{\includegraphics[width=\linewidth]{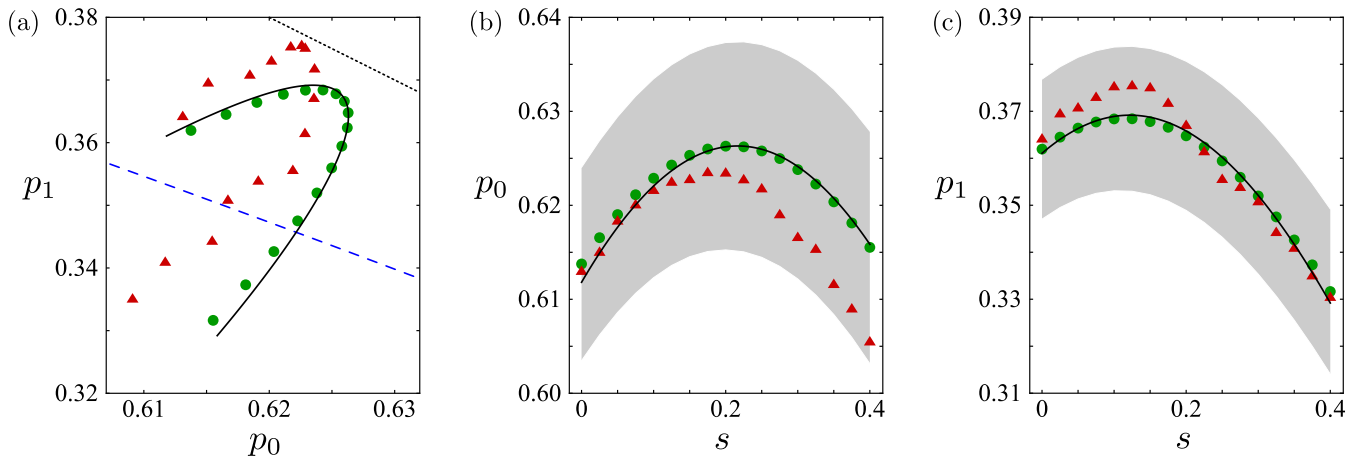}}
\caption{Theoretical fit (solid black line) to the probabilities $p_0(s)$ and $p_1(s)$ estimated from experimental data using pattern functions (green circles). 
Panel (a) depicts the probability pairs in the $p_0,p_1$ plane while panels (b) and (c) show the dependence of $p_0(s)$
and $p_1(s)$ on $s$. The grey areas in panels (b,c) indicate one standard deviation of estimates obtained by pattern functions.
The red triangles represent results of maximum likelihood estimation of $p_0(s)$ and $p_1(s)$ from the experimental data. The blue dashed line in panel (a) indicates the maximum $p_1$
achievable by Gaussian states and their mixtures for a given $p_0$ while the dotted black line marks the ultimate physical boundary $p_0+p_1=1$.}
\label{datafitfig}
\end{figure*}

Next we compare the experimental results with the theoretical model developed in Sec. IV. 
We optimized the model parameters to obtain the best fit to the experimental data shown in Fig. \ref{resultsfig}.
In the numerical optimization we fixed the transmittance of the tap-off beam splitter, $R=0.077$,
and the homodyne detection efficiency $\eta_H=0.80$, which were both reliably determined by independent measurements. 
We also fixed the single-photon efficiency $\eta=0.08$. Although $\eta$ is estimated with a large uncertainty, its value
 mainly influences the success probability of the experiment while the shape of the theoretical fit remains practically unchanged when $\eta$ varies
 within the uncertainty interval.
The fitting  yields the following parameters: $V_x=0.364$, $V_p=0.705$, $Q=0.625$, and $n_{\mathrm{th}}=0$.
This theoretical fit is compared with experimental data in Fig.~\ref{datafitfig} and we achieve very good agreement between theory and experiment. 
For comparison we also plot in Fig.~\ref{datafitfig} the results of maximum likelihood estimation of $p_0(s)$ and $p_1(s)$. 
The ML estimates qualitatively agree with the results of linear reconstruction. The maximum likelihood method
provides slightly higher estimates of $p_1$ and slightly lower estimates of $p_0$  than linear reconstruction but the 
difference is less than one standard deviation, c.f. Fig.~\ref{datafitfig}(b,c).

\section{Conclusions}

In summary, we have experimentally demonstrated the quantum non-Gaussian character of a noisy single-photon squeezed state 
generated by conditional photon subtraction from pulsed squeezed vacuum. Our approach based on a 
witness constructed from probabilities of squeezed vacuum and single-photon states allowed us to reveal the quantum non-Gaussian character of the state
with reasonably high statistical confidence. This was achieved  by an optimization of the variable squeezing parameter $s$.
The generalized witness $W(a,s)$ can identify a much broader class of quantum non-Gaussian states than the original witness $W(a)$
based on the photon number probabilities $p_0$ and $p_1$ \cite{Filip11,Jezek11}
because optimization over various Gaussian operations allows one to suitably match the witness to a given state. 
We emphasize that the witness is able to detect the quantum non-Gaussian character of states with positive Wigner function. 

We have shown that the pattern functions represent an efficient method of estimation of the witness $W(a,s)$ 
from the experimental homodyne data. Moreover, this approach provides reliable easy-to-calculate estimates of the statistical error bars. 
The probability $p_1(s)$ is an overlap of the measured state $\rho$ with squeezed single-photon state.
Since this latter state  contains only odd Fock states in its Fock-state expansion, the probability $p_1(s)$ provides a lower bound 
on the negativity of the Wigner function at the origin of phase space.
In particular, if $p_1(s)>0.5$ for any $s$, then the Wigner function is negative at the origin. 
The pattern functions derived in the present work thus also provide a useful tool for probing the negativity of the Wigner function without 
the need for a full reconstruction of photon number distribution. A detailed analysis of this  will be the subject of a future work.

\acknowledgments

This work was supported by the Czech Science Foundation (P205/12/0577) and the Danish Research Agency
(Project No. FNU 09-072623).


\begin{thebibliography}{99}
\bibitem{anote}
If $a\geq 1$ then $ap_0+p_1$ is trivially maximized by a Gaussian vacuum state with $p_0=1$ and the witness becomes useless.
\end{thebibliography}
\end{document}